\documentclass[%
 reprint,
superscriptaddress,
nofootinbib,
 amsmath,amssymb,
 aps,
]{revtex4-2}
\usepackage{graphicx}
\usepackage{dcolumn}
\usepackage{bm}
\usepackage[version=4]{mhchem} 
\usepackage{subfig}
\usepackage{float}
\usepackage{caption}
\usepackage{multirow}
\usepackage[colorlinks,
            linkcolor=red,
            anchorcolor=blue,
            citecolor=blue
           ]{hyperref}

\graphicspath{ {figure/} }
\begin{document}

\title{Cosmic-ray deuteron excess from a primary component}

\author{Xing-Jian Lv}
\email{lvxj@ihep.ac.cn}
 \affiliation{%
 Key Laboratory of Particle Astrophysics, Institute of High Energy Physics, Chinese Academy of Sciences, Beijing 100049, China}
\affiliation{
 University of Chinese Academy of Sciences, Beijing 100049, China 
}%
 \author{Xiao-Jun Bi}
 \email{bixj@ihep.ac.cn}
\affiliation{%
 Key Laboratory of Particle Astrophysics, Institute of High Energy Physics, Chinese Academy of Sciences, Beijing 100049, China}
\affiliation{
 University of Chinese Academy of Sciences, Beijing 100049, China 
}%
\author{Kun Fang}
\email{fangkun@ihep.ac.cn}
\affiliation{%
 Key Laboratory of Particle Astrophysics, Institute of High Energy Physics, Chinese Academy of Sciences, Beijing 100049, China}
 \author{Peng-Fei Yin}
\email{yinpf@ihep.ac.cn}
\affiliation{%
 Key Laboratory of Particle Astrophysics, Institute of High Energy Physics, Chinese Academy of Sciences, Beijing 100049, China}
\author{Meng-Jie Zhao}
\email{zhaomj@ihep.ac.cn}
 \affiliation{%
 Key Laboratory of Particle Astrophysics, Institute of High Energy Physics, Chinese Academy of Sciences, Beijing 100049, China}
\affiliation{
China Center of Advanced Science and Technology, Beijing 100190, China 
}%



\date{\today}

\begin{abstract}
The recent AMS-02 measurements of cosmic-ray (CR) deuteron flux suggest the presence of primary deuterons in quantities far exceeding predictions from Big Bang nucleosynthesis. This poses a significant challenge to modern astrophysics, as no known processes can account for such large amounts of deuterons without violating existing constraints~\cite{Epstein:1976hq}. In contrast, it has recently been proposed that the AMS-02 measurements can be explained by a purely secondary origin when contributions from heavier nuclei are considered. In this study, we recalculate the secondary deuteron flux using production cross sections updated with the latest data. We find that the deuteron production cross sections are overestimated in the commonly used calculation tools for CR propagation, and a primary deuteron component is still necessary. We then propose a novel process for generating primary deuterons at CR sources through a fusion mechanism, which is naturally unique to deuterons. This model could explain the observed deuteron excess while maintaining consistency with other CR measurements.
\end{abstract}
\maketitle


\section{\label{sec:level1}INTRODUCTION}
Big Bang nucleosynthesis, a cornerstone of the standard model of cosmology, predicts the formation of deuterium shortly after the Big Bang.  During this epoch, temperatures had declined sufficiently to allow for the fusion of protons with neutrons, yet remained high enough to prevent further fusion of deuterium into $\ce{^{4}He}$. This fleeting window resulted in a predicted cosmic abundance of deuterium at $D/H \approx 3 \times 10^{-5}$, which aligns well with observational data (e.g., Ref.~\cite{Cyburt:2015mya}).

Interestingly, the deuterons\footnote{The ionized nucleus of deuterium.} are far more abundant within cosmic rays (CRs), comprising approximately 2\% to 3\% of the proton abundance. This observation initially appears paradoxical, as deuterium is expected to be depleted in stars, with no known mechanisms capable of reversing its decline to such a degree~\cite{Prodanovic:2003bn}. However, the theory of galactic CR propagation provides a natural explanation: the observed deuterons are produced through fragmentation processes involving more abundant primary CR nuclei, which are trapped in the Galactic magnetic field for millions of years~\cite{Wang_2002, PAMELA:2015kyy}. It has long been assumed that nearly all CR deuterons arise from collisions between $\ce{^{4}He}$ ions and nuclei in the interstellar medium (ISM), a phenomenon that also generates CR $\ce{^{3}He}$. Consequently, it is expected that the CR spectra of deuterons and $\ce{^{3}He}$ should exhibit similar energy dependencies~\cite{Coste:2011jc, Gomez-Coral:2023tib}. 

Recently, the AMS-02 collaboration published the precise measurement of the  energy spectrum of deuterons, based on data from 21 million deuteron nuclei within the rigidity range of 1.9 to 21 GV, collected between May 2011 and April 2021~\cite{AMS:2024idr}. Notably, the measured deuteron flux is significantly higher than the secondary deuteron flux predicted by the CR propagation model above 5 GV. Moreover, there is a conspicuous misalignment in the energy dependences between the deuteron and $\ce{^{3}He}$ spectra up to $20$~GV. The AMS-02 collaboration has interpreted these novel phenomena as evidence for the presence of a primary-like deuteron component~\cite{AMS:2024idr}. 

Subsequently, Ref.~\cite{Yuan:2024emf} highlighted the challenges of introducing a significant primary component of deuterons. Instead, the authors proposed that the observed deuteron flux could still be consistent with a secondary origin if the contributions from heavy incident nuclei with $Z > 8$ are taken into account. 
However, it is imperative to carefully scrutinize the production cross sections of deuterons in the evaluation of secondary CR fluxes. As we shall demonstrate later, the deuteron production cross sections employed by the commonly used calculation tools for CR propagation are overestimated in comparison to some of the latest measurements, particularly concerning heavy incident nuclei. Therefore, a purely secondary origin still struggles to fully explain the AMS-02 results.

Consequently, the latest AMS-02 deuteron flux measurements necessitate a primary contribution. However, the production of substantial quantities of non-primordial deuterons remains a conundrum with no recognized explanation in the literature. In this study, we propose a potential origin for primary\footnote{We use \textit{primary} to refer to those secondary components that are not produced through interactions with the ISM during the CR propagation.} CR deuterons. We assume that the spectral hardenings observed in recent measurements of CR secondary-to-primary ratios by AMS-02 and, notably, by DAMPE at $\sim$100 GeV/n~\cite{AMS:2021nhj, Collaboration:2022vwu}, may be attributed to the production and acceleration of secondary particles at the source sites. In this scenario, deuterons produced by proton-proton fusion, which is a process unique to deuterons as the lightest composite nucleus, could account for the peculiar deuteron excess.

The paper is organized as follows. We revisit the deuteron excess after modifying its fragmentation cross sections in Section~\ref{sec:cross section}, followed by the presentation of our model for primary CR deuterons in Section~\ref{sec:method}. The results and discussions are detailed in Section~\ref{sec:results}, and conclusions are provided in Section~\ref{sec:conclusion}.

\section{\label{sec:cross section} The Deuteron Excess Revisited}
As introduced in Section~\ref{sec:level1}, Ref.~\cite{Yuan:2024emf} has clearly illustrated the importance of including nuclei with $Z >8$ when calculating the contributions to secondary deuterons. For the production cross section into deuterons, which is crucial for accurately determining the yield fluxes of secondary particles, they utilized the parameterizations of deuteron production cross sections from Ref.~\cite{Coste:2011jc} (hereafter CDMP12). The approach of CDMP12 for the fragmentation cross sections of nuclei with $A > 4$ is based on the following factorization:
\begin{equation}
\begin{aligned}\label{xs factorization}
\sigma^{\mathrm{P}p \rightarrow \ce{F}}\left(E_{k / n}, A_{\mathrm{P}}\right)=&\gamma_{\mathrm{P}}^{\ce{F}} \times f\left(E_{k / n}, A_{\mathrm{P}}\right) \\
&\times\sigma_{\text {breakup }}^{\ce{^{4}He}p\to \ce{^{3}He}}\left(E_{k / n}\right)\;,
\end{aligned}
\end{equation}
where $\sigma^{\mathrm{P}p \rightarrow \ce{F}}$ is the fragmentation cross section for the projectile P incident upon a target proton producing the fragment F, $A_{\mathrm{P}}$ is the mass number of the projectile, $f(E_{k / n}, A_{\mathrm{P}})$ is given by Eq.~(B.2) of CDMP12, $\sigma_{\text {breakup }}^{\ce{^{4}He}p\to \ce{^{3}He}}$ is the breakup cross section of $\ce{^{4}He}$ colliding on a proton to yield $\ce{^{3}He}$, and $\gamma_{\mathrm{P}}^{\ce{F}}$ is an energy-independent factor determined from measured data points.
\begin{figure}[htbp]
\includegraphics[width=0.48\textwidth]{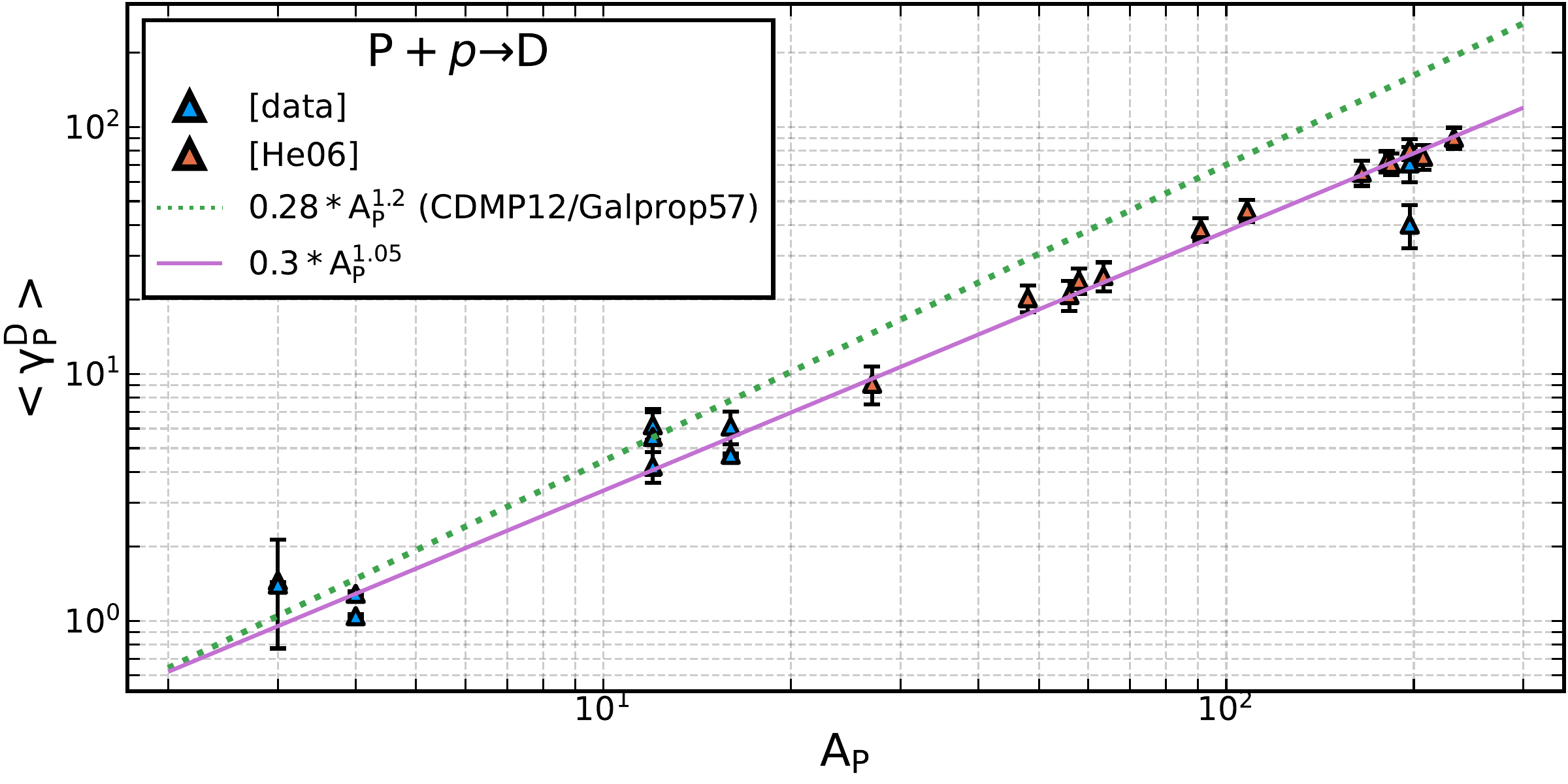}\\
\captionsetup{justification=raggedright}
    \caption{The revised $\gamma_{\mathrm{P}}^{\ce{D}}$ used in this work (solid line) compared with that of CDMP12~\cite{Coste:2011jc}(dotted line). These points are calculated according to the factorization of Eq.~(\ref{xs factorization}).
    Points labeled as [He06] are from Ref.~\cite{Herbach:2006rw}. Points labeled as [data] are from Ref.~\cite{Bazarov:2005qu,Enke:1999vwa,Letourneau:2002fy,Korejwo:2000pf} and the data compilation provided by CDMP12.}
\label{fig:cross sections}
\end{figure}

In determining the form of $\gamma_{\mathrm{P}}^{\ce{D}}$, CDMP12 did not include the cross-section measurements of incident nuclei heavier than oxygen, and provided $\gamma_{\mathrm{P}}^{\ce{D}} = 0.28A_{\mathrm{P}}^{2.1}$. Ref.~\cite{Herbach:2006rw} measured the cross section of deuterons produced in proton-induced fragmentation reactions with targets ranging from Al to Th. After including these data points, we find that CDMP12's $\gamma_{\mathrm{P}}^{\ce{D}}$ factor is too high for heavy nuclei. Consequently, we introduce a refined factor form as follows:
\begin{equation}\label{gamma D}
    \gamma_{\mathrm{P}}^{\ce{D}} = 0.3A_{\mathrm{P}}^{1.05}\;.
\end{equation}
A comparison of $\gamma_{\mathrm{P}}^{\ce{D}}$ factor between CDMP12's and our proposed heavy-nuclei revision is shown in Fig.~\ref{fig:cross sections} alongside the measurements. It is evident that our revision provides a better fit to the data, and would predict less deuteron production. Additionally, we also find that CDMP12 overestimated the fragmentation cross sections for $\ce{^{3}He}$ from heavy nuclei compared with the measurements from Ref.~\cite{Herbach:2006rw}. However, this issue lies outside the scope of our current work and will be addressed in future studies.

As of now, there has been no direct measurement of the deuteron cross section for $\ce{^4He}$-target channels. To determine these cross-sections, we conduct a two-step analysis. In the first step, we calculate some deuteron cross sections for $\ce{^4He}$-target channels with specific projectiles, using data obtained from other targets. Specifically, we estimate the deuteron production cross sections of $\ce{^4He}$, $\ce{^{12}C}$, and $\ce{^{16}O}$ projectiles upon the $\ce{^4He}$-target based on measurements of [Li75]~\cite{osti_971496} and [Ab81]~\cite{SKM-200:1980bbu}. Subsequently, we extrapolate deuteron cross sections for other $\ce{^4He}$-target channels through interpolation. Further details on this analysis can be found in Appendix~\ref{sec: heXS}.

As illustrated in Fig.~\ref{fig:he4target}, the three derived cross sections can be effectively represented by a simple relation $\sigma=\rm0.6A_\mathrm{P}^{1.05}\sigma^{^4He+\mathit{p}\to^3He}$, which scales Eq.~(\ref{xs factorization}) by a factor of 2 (compared with Eq.~(\ref{gamma D})).
{\footnotesize GALPROP} v57~\cite{Porter:2021tlr} calculates $\sigma$ by scaling Eq.~(\ref{xs factorization}) with a factor of $A_\mathrm{P}^{0.31}$, while {\footnotesize USINE}~\cite{Maurin:2018rmm} scales it with a factor of $4^{0.31}$. 
In Fig.~\ref{fig:he4target}, the predictions from {\footnotesize USINE} are systematically lower than the derived cross section points, while {\footnotesize GALPROP} tends to provide higher predictions for projectiles heavier than $\ce{^{12}C}$ compared to our results.
The $\ce{^4He} + \ce{^4He}$ channel, which contributes the most to the deuteron flux among all $\ce{^4He}$-target channels, exhibits a significant uncertainty and requires precise measurements for improved predictions in the future.
Overall, our prediction of the deuteron flux from $\ce{P} + \ce{^4He}$ channels shows only a slight increase compared to GALPROP's prediction, representing a minor adjustment relative to the corrections made to the heavy nuclei fragmentation cross sections of $\ce{P} + p$ channels.

\begin{figure}[htbp]
\includegraphics[width=0.48\textwidth]{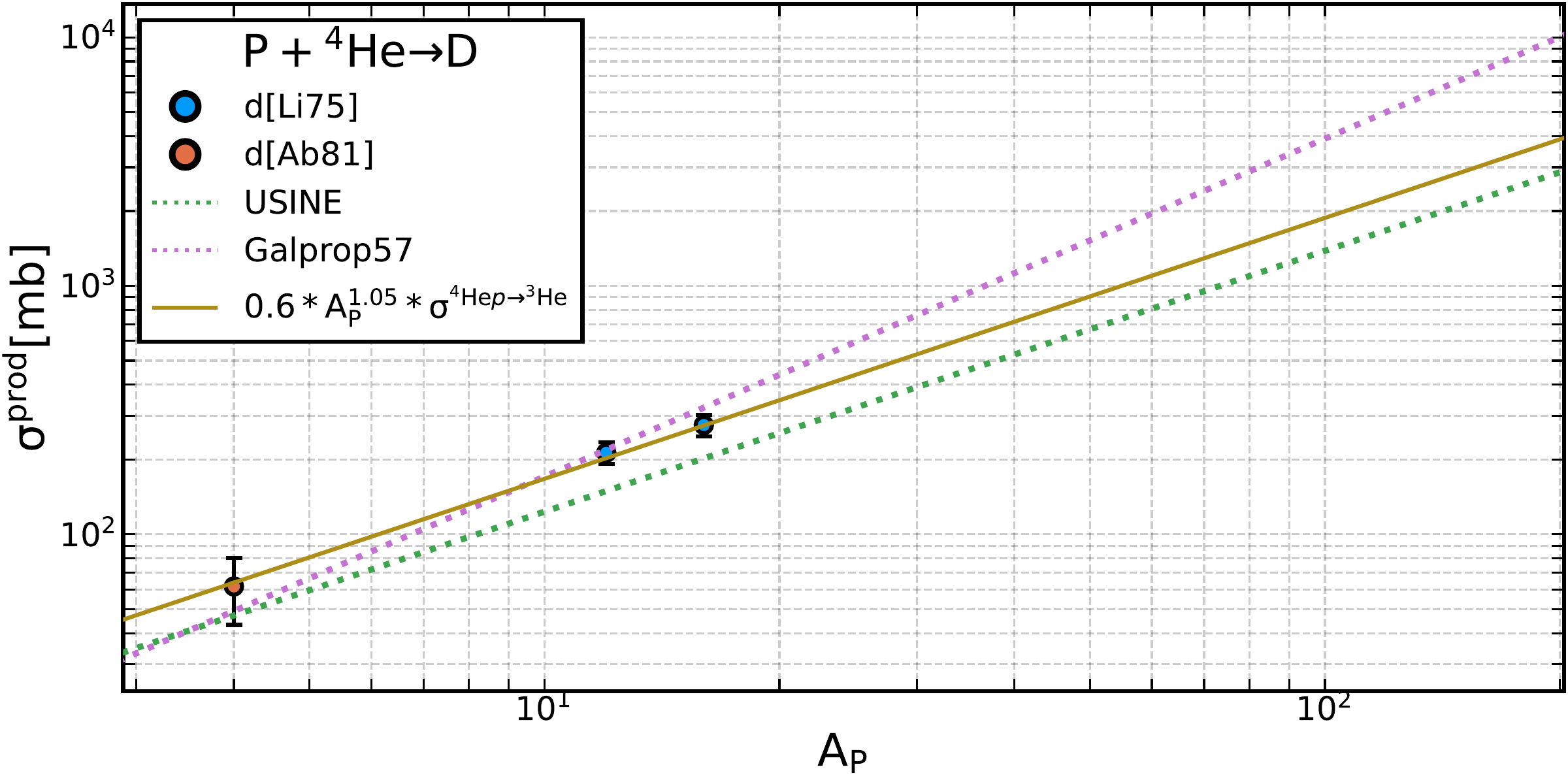}\\
\captionsetup{justification=raggedright}
\caption{{The production cross sections of deuterons from the various projectiles on the $^4$He target at high energies. The derived data points are based on Ref.~\cite{osti_971496, SKM-200:1980bbu}. The purple and green dotted lines are parametrizations used in {\footnotesize GALPROP} v57~\cite{Porter:2021tlr} and {\footnotesize USINE}~\cite{Maurin:2018rmm}. The brown solid line is the fitted result derived by this study.} \label{fig:he4target}}
\end{figure}

\begin{figure}[htbp]
\includegraphics[width=0.48\textwidth]{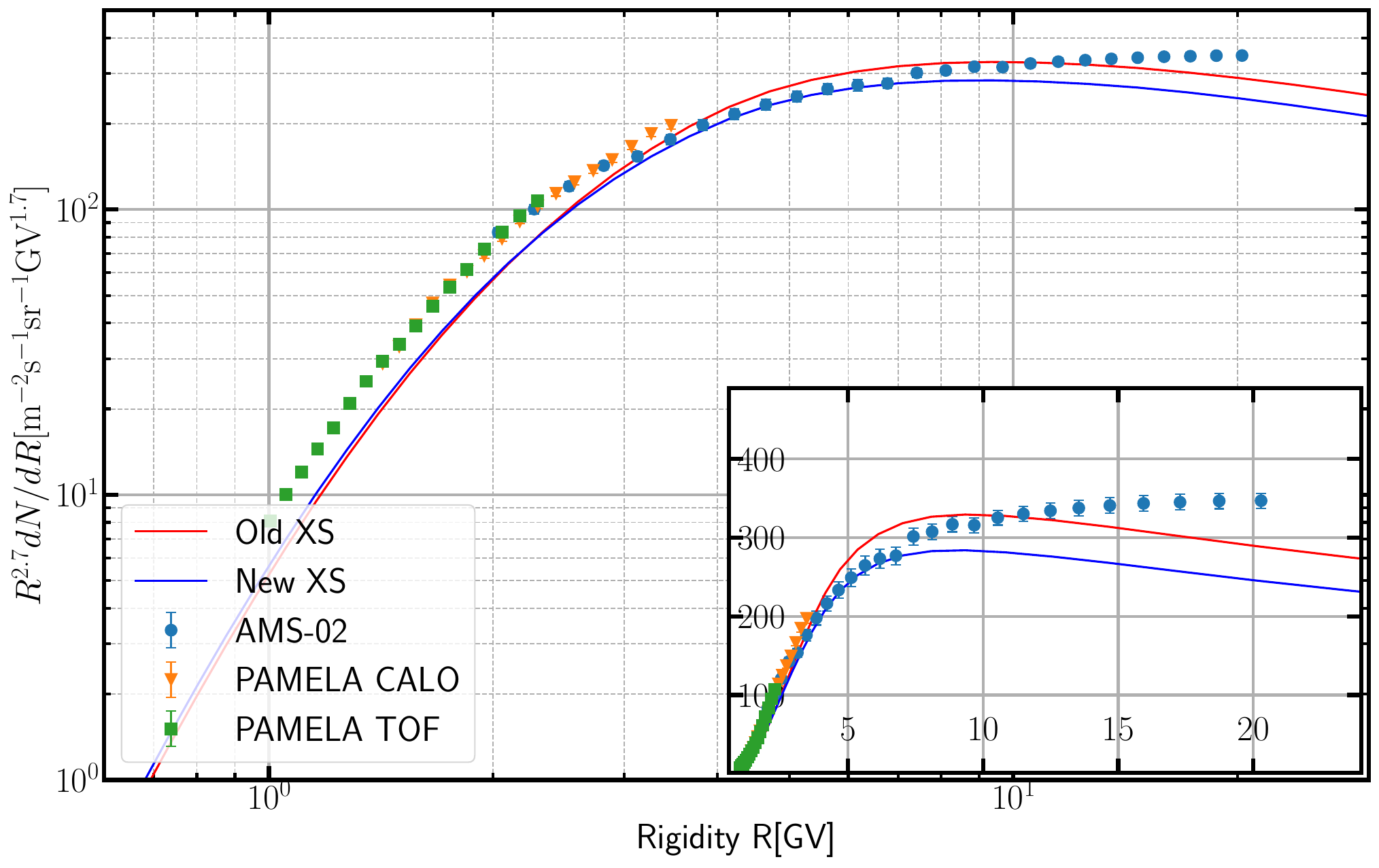}\\
\captionsetup{justification=raggedright}
\caption{Predicted fluxes of pure secondary deuterons using the previous and the updated cross sections compared with measurements~\cite{AMS:2024idr, PAMELA:2018weq}. The inset displays the same plot in linear scale for a clear view of the discrepancy. The result derived with the previous cross sections is similar to that obtained by Ref.~\cite{Yuan:2024emf}. \label{fig:yuan}}
\end{figure}

The modification of fragmentation cross sections results in a reduction in the deuteron flux, of approximately 15\% around 10 GV when adjusting heavy-nuclei contributions, and an increment of about 1.5\% when additionally adjusting $\ce{^4He}$-target contributions.
Due to the poor constraints of the related measurements, non-negligible uncertainties of around 10\% in the secondary deuteron production cross sections should be considered in the analysis, as would be mentioned in Section~\ref{sec:results}.
We illustrate the difference of predicted deuteron flux between the previous and the updated cross sections in Fig.~\ref{fig:yuan}. The excess at the highest energy points appears modest with the previous cross sections, but the deficit becomes significant with the adoption of the revised cross sections at $\sim10$~GV. In addition, even with the utilization of the previous cross sections, the rigidity dependence of the predicted secondary deuterons remains harder than that observed in the AMS-02 data. This discrepancy is clearly delineated in the inset of Fig.~\ref{fig:yuan}, where linear scales are used to enhance clarity. In conclusion, our analysis suggests that the AMS-02 deuteron flux cannot be fully explained by a purely secondary origin.

\section{\label{sec:method} Contributions to Deuteron Flux}
\subsection{Primary Deuteron}
The recent measurements of CR secondary-to-primary ratios by AMS-02 and DAMPE reveal significant spectral hardenings around 100 GeV/n~\cite{AMS:2021nhj, Collaboration:2022vwu}. This phenomenon presents a challenge to conventional production and propagation models of CRs. These models typically posit that secondary particles are produced via the fragmentation of primary CRs interacting with ISM during their propagation through the Milky Way~\cite{Strong:2007nh}. One plausible explanation for these spectral hardenings is the occurrence of similar particle interactions near CR sources, especially in regions where dense molecular clouds surround these sources. Such interactions could give rise to the production of primary boron nuclei, consequently leading to the observed hardenings in the  B/C and B/O ratios~\cite{Zhang:2021xri, Zhang:2022rsx, Ma:2022iji}. This scenario could also account for the observed positron excess~\cite{Fujita:2009wk, Kohri:2015mga, Yang:2018nhs} and the excess in ultrahigh-energy diffuse $\gamma$-ray emission~\cite{TibetASgamma:2021tpz, Zhang:2021xri, Sun:2023ibg}.

\begin{figure}[htbp]
\includegraphics[width=0.48\textwidth]{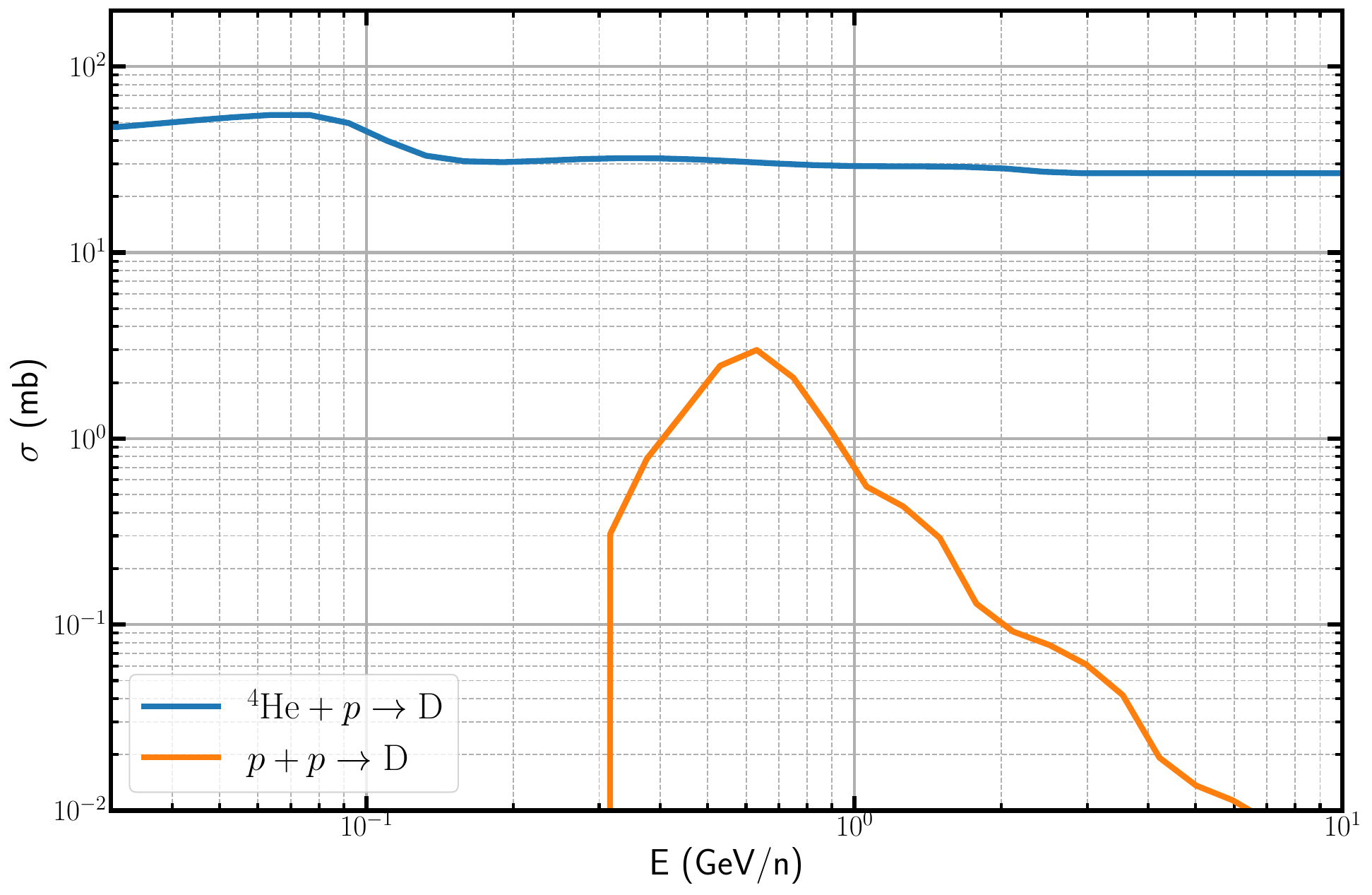}\\
\captionsetup{justification=raggedright}
\caption{The cross sections of two deuteron production channels provided in Ref.~\cite{Coste:2011jc,Porter:2021tlr}: one from the fragmentation of $\ce{^4He}$, and the other from the fusion of protons. \label{fig:ppXS}}
\end{figure}

\begin{figure*}[t]
    \begin{center}
        \subfloat{\includegraphics[width=0.45\textwidth]{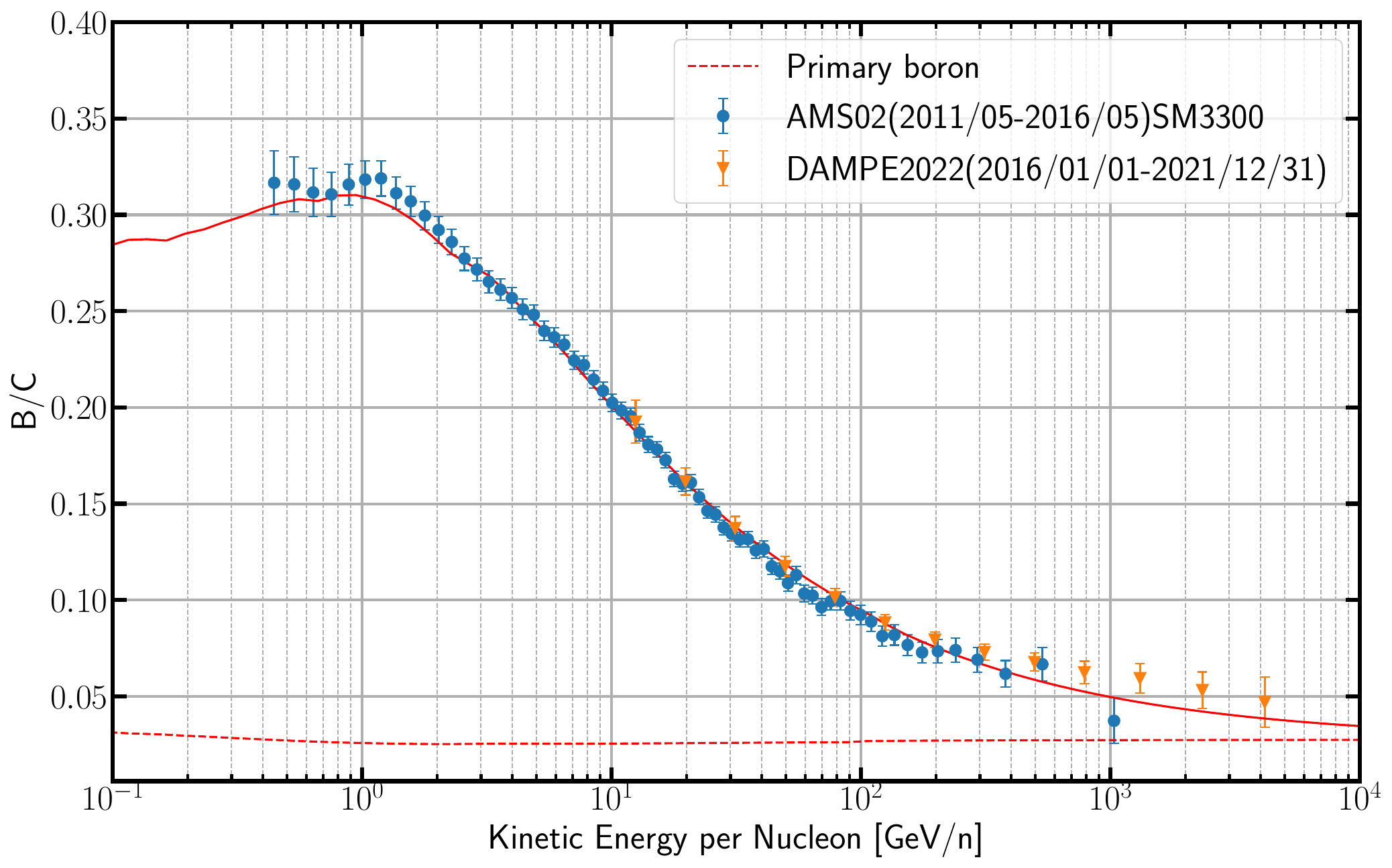}} \hskip 0.03\textwidth
        \subfloat{\includegraphics[width=0.45\textwidth]{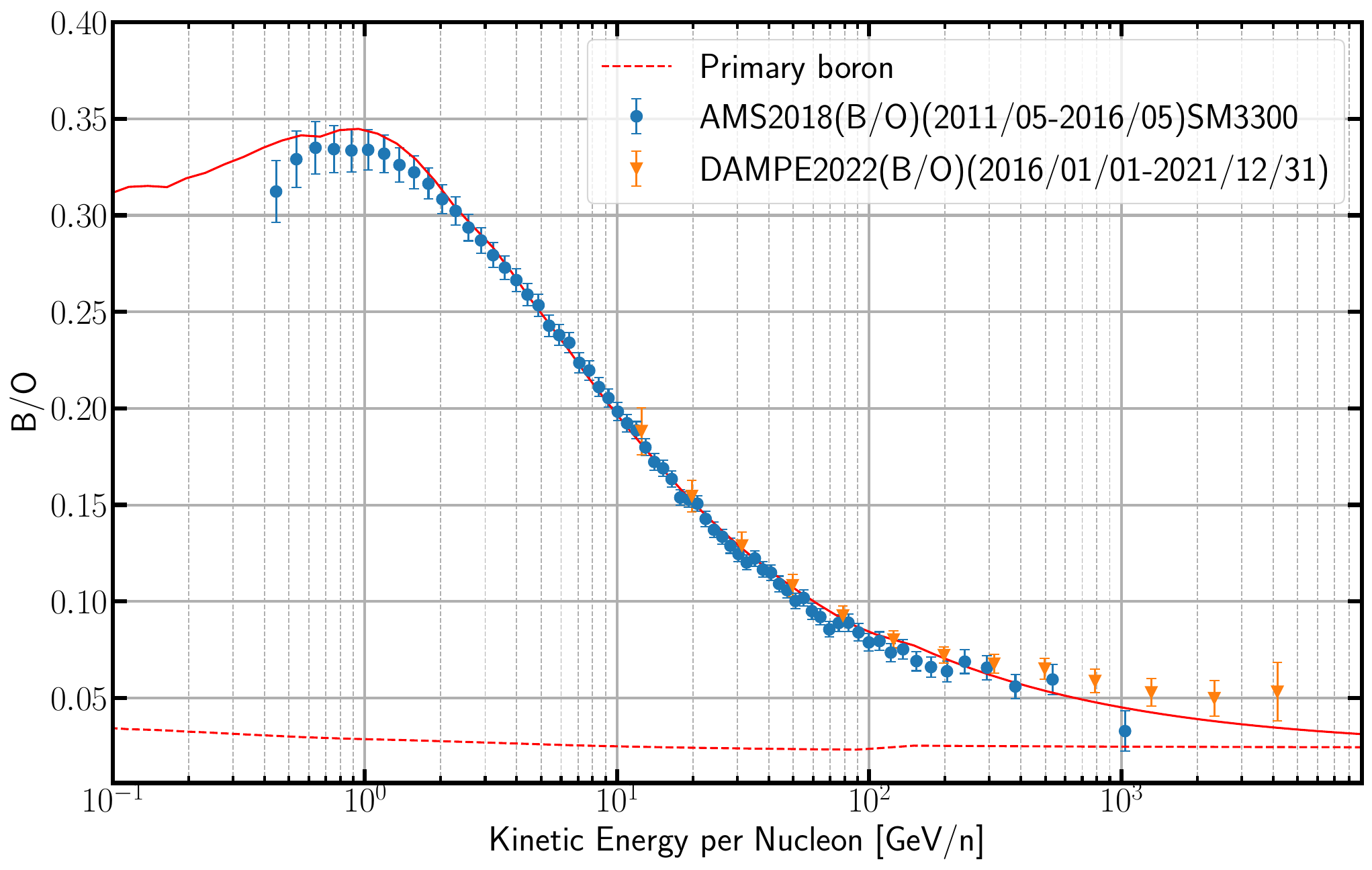}}
    \end{center}
    \captionsetup{justification=raggedright}
    \caption{Best-fit results for B/C, B/O ratios, compared with measurements~\cite{AMS:2018tbl, Collaboration:2022vwu}. In both panels, the solid line represents total boron, while the dashed line indicates contributions from primary boron.}\label{fig:bc}
\end{figure*}

The primary boron produced at sources through fragmentation is assumed to share the same injection spectrum as carbon and oxygen.
As shown in Fig.~\ref{fig:bc}, we determine the normalization of primary boron by adding the secondary boron flux to fit the overall B/C and B/O ratios. 
Notably, the production of primary boron through CR-gas interactions is always accompanied by the production of primary deuterons. Therefore, by utilizing their respective parent particle abundances and production cross sections as a scaling factor, we can calculate the primary deuteron flux from fragmentation of helium associated with the primary boron flux as follows:
\begin{equation}
\begin{aligned}
  & \Phi^{\text{LIS}}_{\text{primary } \ce{^{4}He}+p\to\ce{D}} = \\
  & \Phi^{\text{LIS}}_{\text{primary B}}\times\frac{\Phi^{\text{LIS}}_{\ce{^{4}He}}\sigma^{\ce{^{4}He}p\to \ce{D}}}{\Phi^{\text{LIS}}_{\ce{C}}  \sigma^{\ce{C}p\to \ce{B}} + \Phi^{\text{LIS}}_{\ce{O}} \sigma^{\ce{O}p\to \ce{B}}}\;, \\
\end{aligned}
\end{equation}
where the label LIS denotes the local interstellar flux under consideration.

As the lightest composite nucleus, the deuteron has a unique production channel unavailable to heavier nuclei such as $\ce{^{3}He}$ or boron, namely, the coalescence of two protons to form a deuteron. Despite the cross section for this process being ten times smaller than that of other channels, the abundance of CR protons being approximately tenfold greater than that of $\ce{^{4}He}$ makes this contribution to the secondary deuteron flux non-negligible~\cite{Coste:2011jc}. This fusion process is also expected to occur near the source regions, following the same logic as fragmentation processes. Unlike fragmentation, this fusion cross section $\sigma_{p+p\to \ce{D}}$ is non-vanishing only within a very narrow energy range of around 600 MeV/n, as illustrated in Fig.~\ref{fig:ppXS}. The contribution from primary fusion deuterons can be calculated as
\begin{equation}
\begin{aligned}
    & \Phi^{\text{fusion}}_{\text{primary } p+p\to\ce{D}} = \\
    & \Phi^{\text{LIS}}_{\text{primary B}}\times\frac{\Phi^{\text{LIS}}_{p}\sigma^{pp\to \ce{D}}}{\Phi^{\text{LIS}}_{\ce{C}}  \sigma^{\ce{C}p\to \ce{B}} + \Phi^{\text{LIS}}_{\ce{O}} \sigma^{\ce{O}p\to \ce{B}}}\;. \\ \label{eqn: fusion}
\end{aligned}
\end{equation}

Despite the fusion process being confined to a narrow energy range around 600 MeV/n,
the fusion deuterons originating in the vicinity of accelerating sources can undergo further acceleration by shocks near these sources~\cite{Berezhko:2003pf, Blasi:2009hv}. Consequently, the resulting spectrum is expected to exhibit a power law behavior akin to that of primary fragmentation deuterons. Accordingly, we reshape the contribution of primary fusion deuterons given by Eq.~(\ref{eqn: fusion}) to the LIS deuteron flux to match the shape of the LIS helium flux, incorporating a low-energy cutoff as 
\begin{equation}
    \Phi^{\text{LIS}}_{\text{primary } p+p\to\ce{D}} \propto \Phi^{\text{LIS}}_{\ce{^{4}He}} \times \exp[-(E_{\text{cut}}/E)^3]\; .
\end{equation}
In the calculation, the integral intensity of deuteron is preserved as in Eq.~(\ref{eqn: fusion}).
The low-energy cutoff is the sole free parameter, as all other parameters are fully constrained by the B/C and B/O ratios. 
We will discuss the justification for introducing this cutoff and examine its impact on the deuteron flux in Section~\ref{sec:results}.

\subsection{Secondary Deuteron}
We use the numerical tool {\footnotesize GALPROP}~\cite{Strong:1998pw} v57\footnote{Current version available at \url{https://galprop.stanford.edu/}.} to calculate
the production and propagation of secondary particles. We adopt the diffusion propagation framework with stochastic reacceleration in the ISM, which has been shown to accurately reproduce secondary-to-primary ratios~\cite{Yuan:2017ozr, Yuan:2018lmc}. We use the production cross sections for secondary deuterons fitted to the latest measurements, as described in Section~\ref{sec:cross section}, and include all nuclei up to $Z = 28$ (nickel) in our calculations. The contribution from the fusion of CR protons with ISM protons is also included in the calculation. For the injection spectra of primary CRs,
we employ a broken power law to account for the hardenings observed in the primary CR spectra. The details regarding the propagation calculations and parameter determination are provided in Appendix~\ref{sec: parameters}.

\section{\label{sec:results} Results \& Discussion}
\begin{figure}[htbp]
\includegraphics[width=0.48\textwidth]{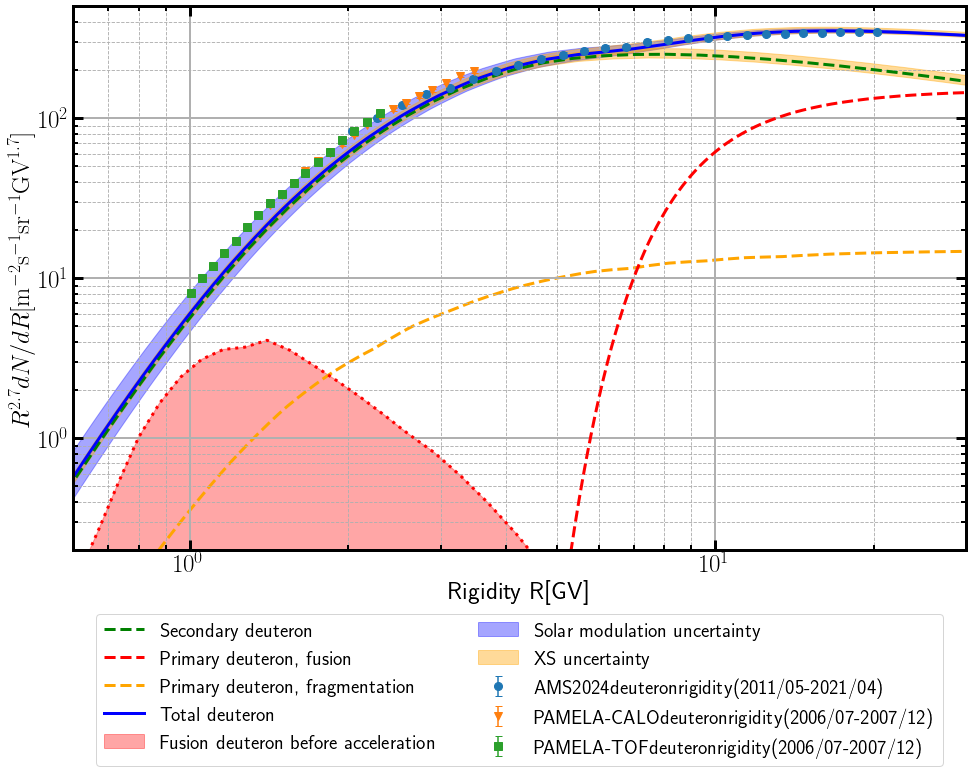}\\
\captionsetup{justification=raggedright}
\caption{Expected deuteron flux in comparison with measurements~\cite{AMS:2024idr, PAMELA:2018weq}. The solid blue line represents the total spectrum at the top of the atmosphere considering a solar modulation potential of $\Phi=600$ MV, with the blue-shaded band indicating the uncertainty range for $\Phi=600\pm 100$ MV. The orange-shaded band represents uncertainties due to secondary deuteron production cross sections. The green, orange, and red dashed lines correspond to the secondary, primary fragmentation, and primary fusion contributions to the deuteron flux, respectively. The red-shaded region represents the primary fusion deuteron spectrum before acceleration. The low-energy cutoff of the accelerated primary fusion spectrum is assumed to be $4$~GeV/n. \label{fig:deuteron}}
\end{figure}

The calculated flux of deuterons is shown in Fig.~\ref{fig:deuteron}, along with the observational data from PAMELA~\cite{PAMELA:2018weq} and AMS-02~\cite{AMS:2024idr}. Within the framework of this study, the deuteron flux includes three major components: i) a secondary component arising from the fragmentation of nuclei up to $Z = 28$ (nickel) during the propagation process; ii) a primary fragmentation component originating from the CR sources; iii) a primary fusion component also originating from the sources. To account for the solar modulation effect on the spectrum inside the heliosphere, we adopt the force-field approximation~\cite{Gleeson:1968zza}. The modulation potential is taken to be 600 MV for all nuclei. A range of $\Phi = 600 \pm 100$ MV is considered to account for variations in data collection periods and the inherent limitations of the force-field approximation~\cite{Potgieter:2013pdj, tomassettiSolarNuclearPhysics2017}, visually represented as a blue-shaded band in the figure. As discussed in Section~\ref{sec:cross section}, there remain non-negligible uncertainties in the secondary deuteron production cross sections, particularly concerning uncertainties of heavier-nuclei and $\ce{^4He}$-target modification ($\sim10\%$ around 10 GV). The range of secondary deuteron flux allowed by the current cross-section data uncertainties is illustrated as an orange-shaded band.  

As shown in Fig.~\ref{fig:deuteron}, the predicted deuteron flux including three components aligns well with the measurements, taking into account the influences of solar modulation and uncertainties in cross section. Notably, the discrepancy between the AMS-02 data and the theoretical prediction at above $\sim 10$~GV can be resolved by introducing a primary fusion component with a low-energy cutoff at $4$~GeV/n. In comparison, the contribution of the primary fragmentation component to the deuteron flux is minor.

\begin{figure}[htbp]
\includegraphics[width=0.48\textwidth]{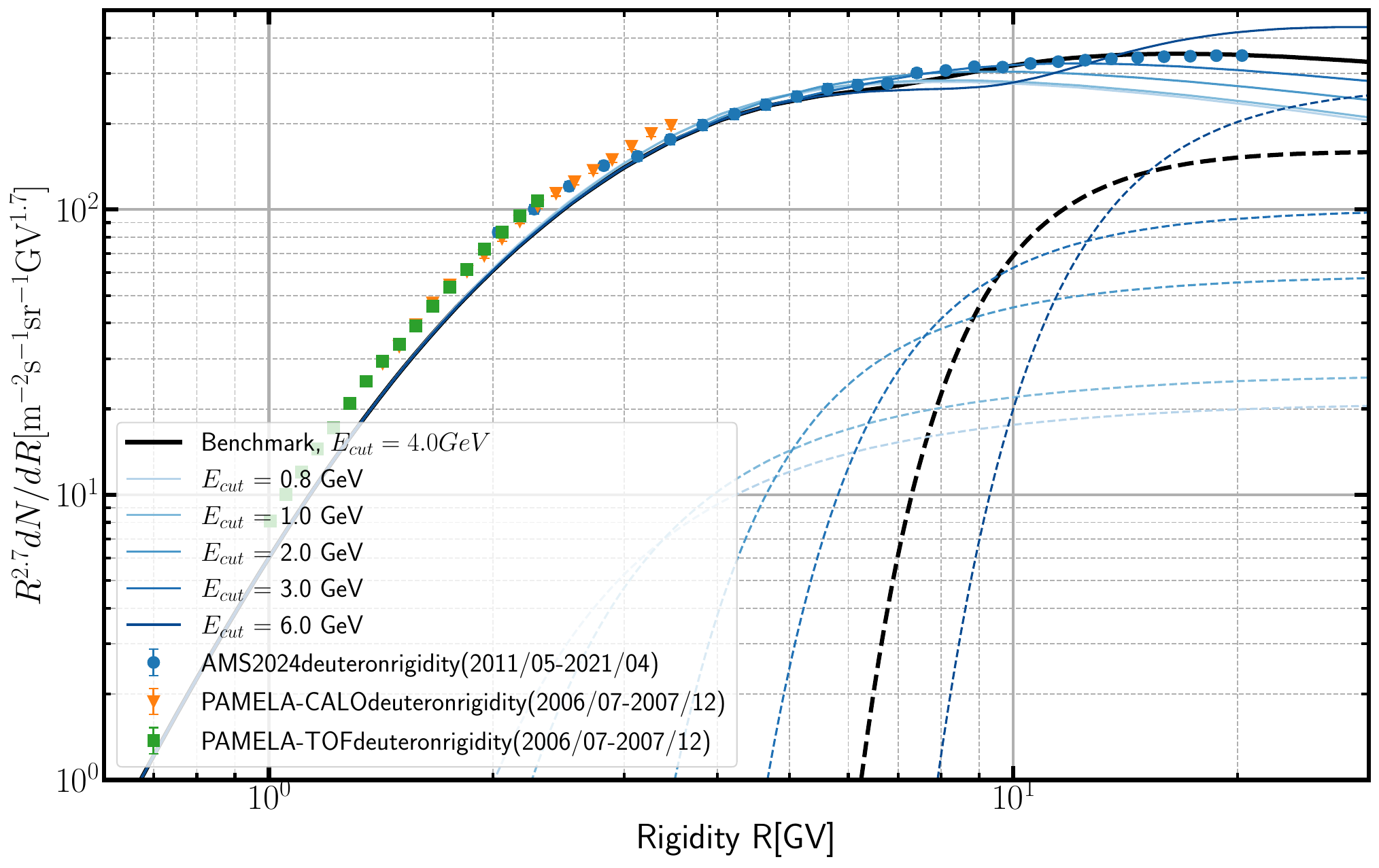}\\
\captionsetup{justification=raggedright}
\caption{Effects of changing the low-energy cutoff for the accelerated primary fusion deuterons. The solid lines represent the total deuteron flux, while the dashed lines indicate contributions from primary fusion deuterons. The references are the same as in Fig.~\ref{fig:deuteron}.\label{fig:deuteron ecut}}
\end{figure}

As the low-energy cutoff of the primary fusion component is the only free parameter in our model for calculating the deuteron flux, we further investigate the impact of varying  $E_{\text{cut}}$. Figure~\ref{fig:deuteron ecut} illustrates the contributions of primary fusion deuterons and the total deuteron fluxes at Earth for several cutoff values -- $0.8$, $1.0$, $2.0$, $3.0$, $6.0$ GeV/n, in addition to the benchmark value of $4.0$ GeV/n. As the cutoff energy decreases, the contribution of primary fusion deuterons to the high-energy end of the AMS-02 data diminishes,  with a predominant flux shift towards lower energies. As a result, the primary fusion component is insufficient to explain the excess at high energies for $E_{\text{cut}}<4$ GeV/n, while the expected spectrum is harder than the actual measurement for $E_{\text{cut}}>4$ GeV/n.

We notice that a significant low-energy break at a few GeV/n in the injection spectrum for primary CR species has already been suggested \cite{Johannesson:2016rlh,Boschini:2020jty,Pan:2023gji} to explain the latest CR measurements. Ref.~\cite{Pan:2023gji} found that for some species, the spectra indices below this break are even smaller than 1. While the broken power-law form is milder than the cutoff feature adopted in this study, it is important to note that the seed spectra before acceleration for those primary CRs follow Maxwellian distributions~\cite{Park:2014lqa}, which provide an ample reservoir of low-energy particles. In contrast, the seed spectrum for primary fusion deuterons has a much sharper cutoff at low energies, as illustrated by the red-shaded region in Fig.~\ref{fig:deuteron}. Hence, a sharp cutoff for the accelerated primary fusion deuterons appears to be a plausible scenario.

\section{\label{sec:conclusion}Conclusion}
In this study, we study the discrepancy between the theoretical predictions and the observed deuteron flux revealed by the AMS-02 results. Our analysis begins with a thorough examination of the secondary deuterons flux. We find that the deuteron production cross sections utilized in commonly used calculation tools for CR propagation are overestimated compared to some latest measurements, especially for channels involving heavy incident particles. Consequently, considering the constraints on CR propagation parameters derived from secondary-to-primary CR flux ratios like B/C and B/O, we find that the flux of secondary CR deuterons is significantly lower than the deuteron spectrum measured by AMS-02 above $\sim 10$~GV. This discrepancy challenges previous findings based on the commonly used deuteron production cross sections and suggests that a substantial primary CR deuteron component is essential to account for the AMS-02 results.

For the possible origin of primary deuterons, we propose an innovative scenario in which deuterons generated through proton-proton fusion in the CR source region could offer a plausible explanation for the intriguing excess observed in the data. The observed hardening of the B/C and B/O spectra at $\sim100$~GeV/n could potentially be attributed to the superposition of secondary boron produced during propagation and primary boron generated at the sources. Building upon this hypothesis, we posit the presence of primary deuterons originating from heavy nuclei fragmentation and proton-proton fusion at the sources. These primary deuterons are subsequently accelerated and may exhibit similar spectral indices as $^4$He at high energies. By estimating the contribution of primary deuterons alongside primary borons, we find that the unique fusion component, which is absent in $^3$He and boron, can naturally serve as a source of the observed excess in the deuteron flux.

To reproduce the AMS-02 data, our analysis indicates the necessity of implementing a low-energy cutoff at $4$~GeV/n for the accelerated fusion deuteron spectrum. This introduction of a cutoff is phenomenological, not derived from first principles, due to the absence of studies on the acceleration mechanisms of the non-thermal seed spectrum. This parameter, serving as the sole free parameter in our model, plays a pivotal role in determining the flux of primary fusion deuterons above $\sim 10$~GV. Current observations have suggested a significant suppression in the injection spectra of primary nuclei below several GV. However, the non-thermal nature of the seed spectrum of primary fusion deuterons before acceleration precludes direct comparisons.  Hence, future investigations focusing on the acceleration mechanisms of this unique seed spectrum are essential to validate the proposed form of the low-energy spectral cutoff discussed in this study.

\acknowledgments
This work is supported by 
the National Key R\&D Program Grants of China under Grant No. 2022YFA1604802, 
the National Natural Science Foundation of China under the Grants No. 12105292, No. 12175248, No. 12393853, and partially supported by the National Natural Science Foundation of China under grant No. 12342502. 

\bibliography{apssamp}
\appendix
\section{Propagation and injection spectra of cosmic rays\label{sec: parameters}}

\begin{figure*}[t]
    \begin{center}
        \subfloat{\includegraphics[width=0.45\textwidth]{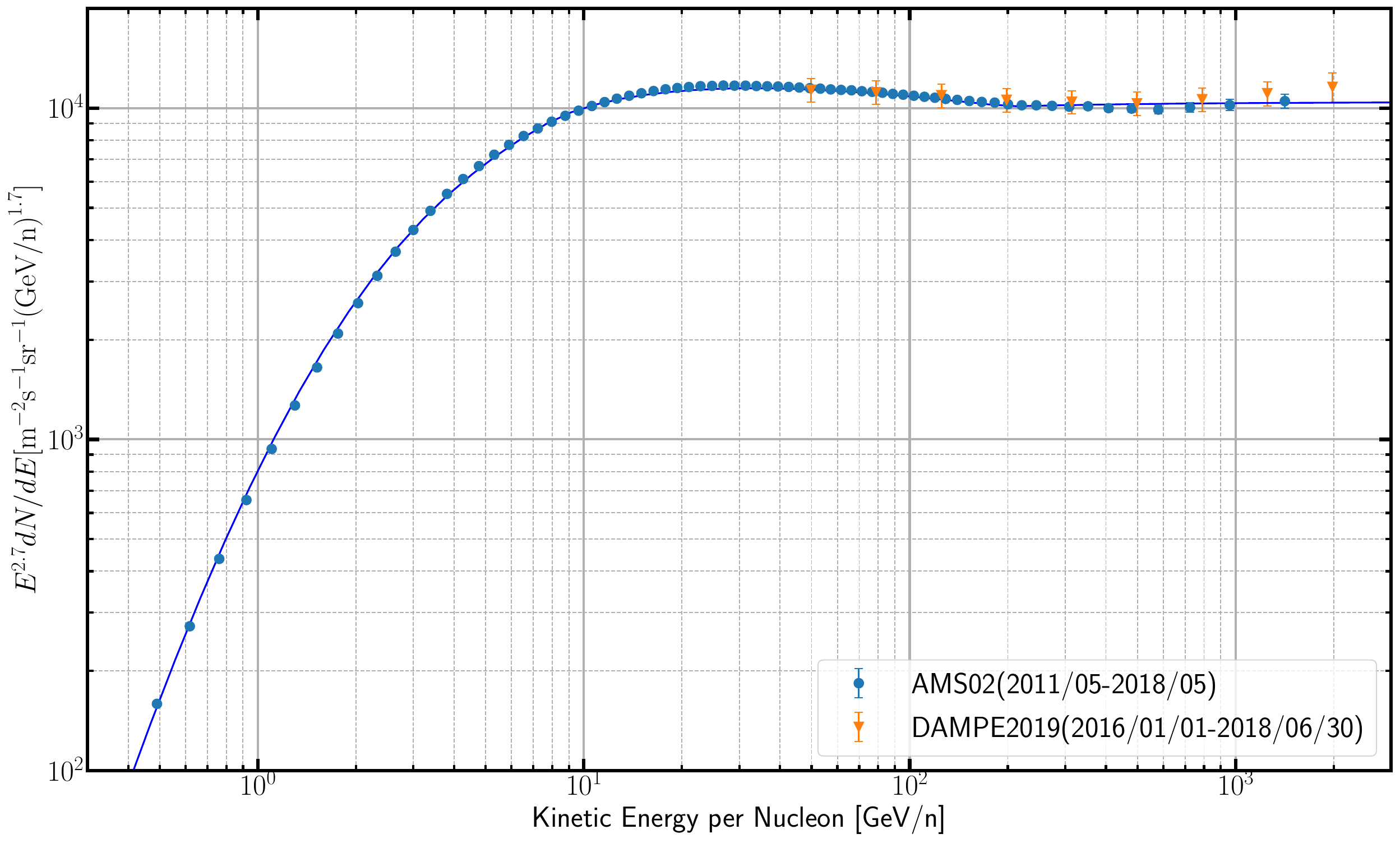}} \hskip 0.03\textwidth
        \subfloat{\includegraphics[width=0.45\textwidth]{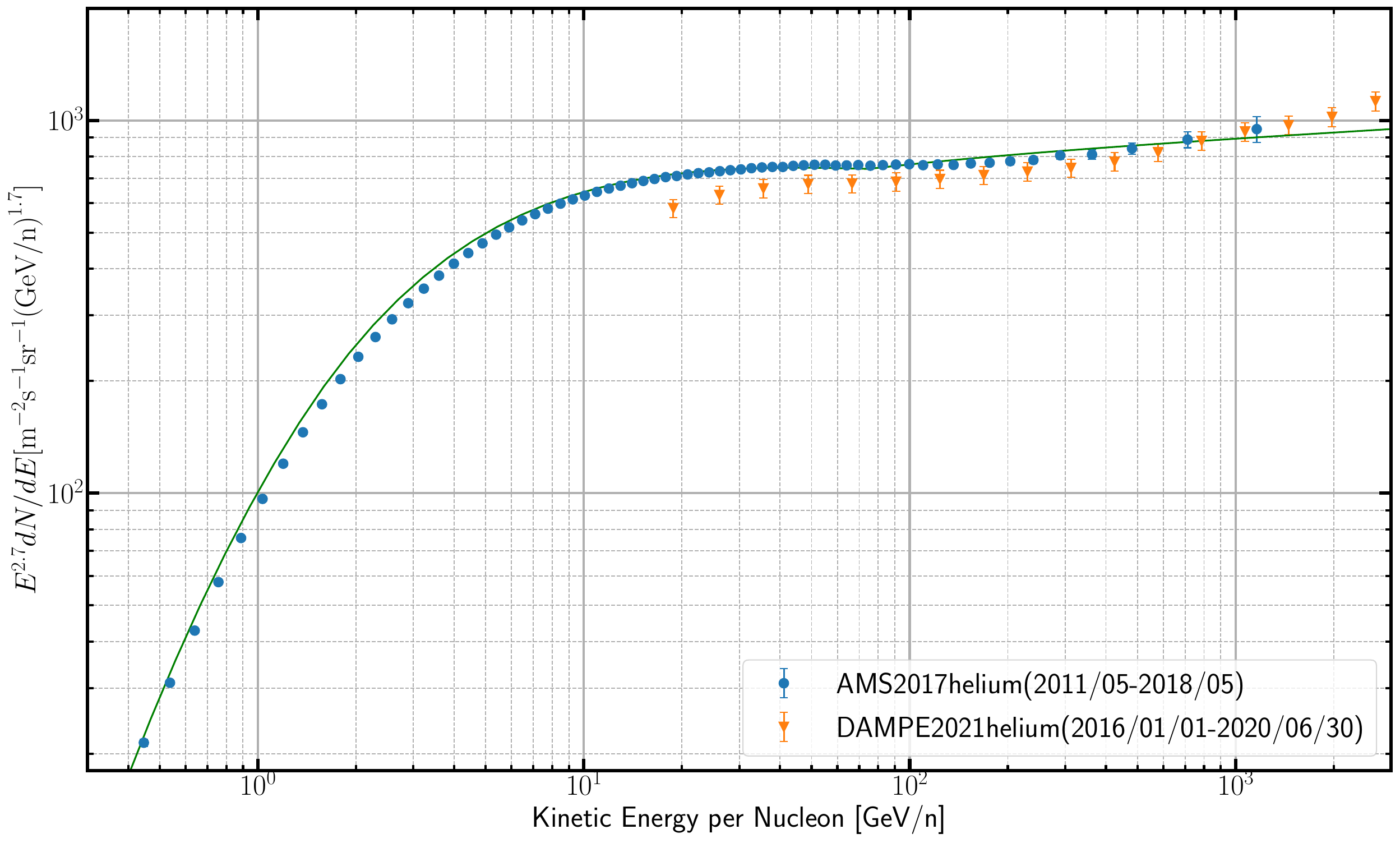}}
    \end{center}
    \captionsetup{justification=raggedright}
    \caption{Best-fit results for proton and helium fluxes, compared with measurements~\cite{AMS:2021nhj, DAMPE:2019gys, Alemanno:2021gpb}}\label{fig:proton}
\end{figure*}

\begin{figure}[htbp]
\includegraphics[width=0.48\textwidth]{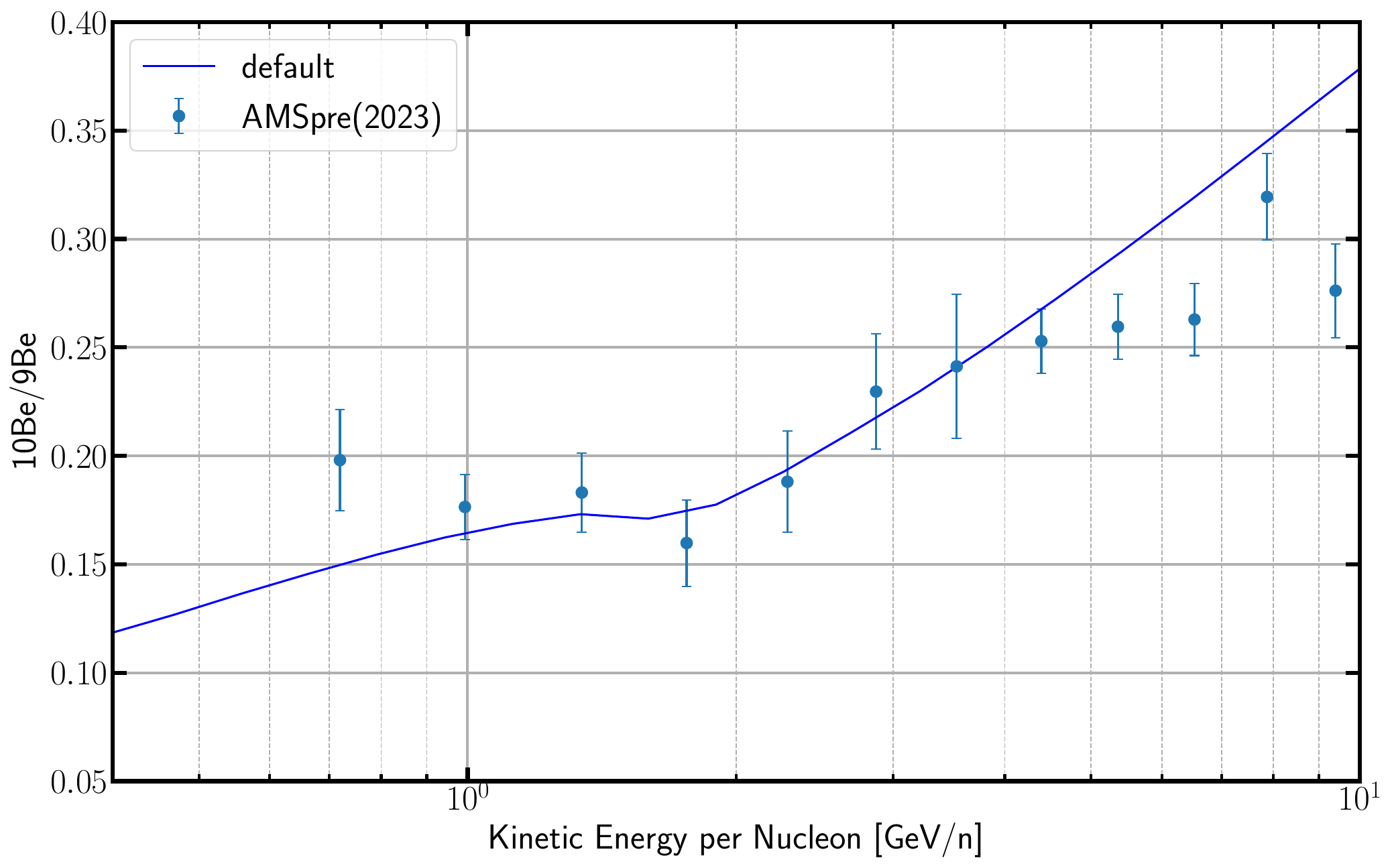}\\
\captionsetup{justification=raggedright}
\caption{Best-fit results for$\ce{^{10}Be}$/$\ce{^{9}Be}$, compared with measurements~\cite{wei2022beryllium}.\label{fig:10be9}}\label{fig:be}
\end{figure}

The propagation equation of Galactic
CRs is expressed as~\cite{Strong:2007nh}:
\begin{equation}
\begin{aligned}
\frac{\partial \psi}{\partial t}= & Q(\mathbf{x}, p)+\nabla \cdot\left(D_{x x} \nabla \psi-\mathbf{V}_c \psi\right)+\frac{\partial}{\partial p}[p^2 D_{p p} \frac{\partial}{\partial p}(\frac{\psi}{p^2})] \\
& -\frac{\partial}{\partial p}[\dot{p} \psi-\frac{p}{3}(\nabla \cdot \mathbf{V}_c) \psi]-\frac{\psi}{\tau_f}-\frac{\psi}{\tau_r}\;,
\end{aligned}
\end{equation}
where $Q(x, p)$ represents the CR source term, $\psi = \psi(x, p, t)$ denotes the CR density per unit momentum $p$ at position $\mathbf{x}$, $ \dot{p} \equiv dp/dt$ is the momentum loss rate, and the time scales $\tau_f$ and $\tau_r$ characterize fragmentation processes and radioactive decays, respectively.  In the framework of diffusive-reacceleration, the momentum space diffusion coefficient, $D_{p p}$ is related to the spatial diffusion coefficient $D_{x x}$ through the relation $D_{p p} D_{x x}=4 p^2 v_A^2/\left(3 \delta\left(4-\delta^2\right)(4-\delta)\right)$~\cite{seoStochasticReaccelerationCosmic1994, berezinskiiAstrophysicsCosmicRays1990a}, where $v_A$ is the Alfv\'en velocity.

The spatial diffusion coefficient is parameterized as a function of rigidity as\footnote{Note that this factor serves as an effective parameterization to describe the intricate process of charged particles traversing turbulent magnetic fields within the Galaxy. In reality, the propagation of charged particles is influenced by the complex structure and dynamics of the Galactic magnetic field, which depends on a diverse spectrum of turbulence spanning various scales, anisotropic diffusion, and localized magnetic irregularities.
\begin{equation}
    D_{xx} = D_0 \, \beta^{\eta} \left(\frac{R}{R_0}\right)^{\delta}\; ,
\end{equation}
where $\beta$ is the particle velocity in units of the speed of light, $\eta$ is an empirical modification of the velocity dependence that allows for a better fit to low-energy data~\cite{DiBernardo:2009ku}, and $D_0$ is the normalization at $R_0 = 4 \, \text{GV}$.}

The injection spectrum is parameterized as:
\begin{equation}
q^i(R) \propto\left\{\begin{array}{cc}
\left(R / R^i_{\mathrm{br0}}\right)^{-\nu^i_0} , & R\leq R^i_{\mathrm{br}0} \\
\left(R / R^i_{\mathrm{br1}}\right)^{-\nu^i_1} , & R^i_{\mathrm{br}0}>R\leq R^i_{\mathrm{br}} \\
\left(R / R^i_{\mathrm{br1}}\right)^{-\nu^i_2} , & R>R^i_{\mathrm{br}}\;\qquad ,
\end{array}\right.
\end{equation}
where $i$ denotes the species of the nuclei. The low-energy break $R_{\mathrm{br}0}$ is introduced to account for the observed low-energy spectral bumps observed in all nuclei~\cite{Johannesson:2016rlh, Phan:2021iht}, while the high-energy break $R_{\mathrm{br}}$ is introduced to account for the hardening observed in all primary nuclei~\cite{AMS:2021nhj}.

For interactions and energy losses, we include all relevant processes for the propagation of nuclei, such as fragmentation, momentum losses, radioactive decay, K capture, ionization losses, and Coulomb losses. 
In order to optimize computational efficiency, processes specific to lepton propagation, like synchrotron radiation, are not considered. 
For inputs such as the source distribution, gas model, and He/H ratio in the interstellar medium, we adopted the default parameters from the ``Pulsar/XCO R-dep" model detailed in the latest GALPROP v57 release~\cite{Porter:2021tlr}.

The data used for the determination of injection and propagation parameters include:
\begin{itemize}
    \item B/C: AMS-02~\cite{AMS:2018tbl}, DAMPE~\cite{Collaboration:2022vwu}
    \item B/O: AMS-02~\cite{AMS:2018tbl}, DAMPE~\cite{Collaboration:2022vwu}
    \item $\ce{^{10}Be}$/$\ce{^{9}Be}$: AMS-02 preliminary~\cite{wei2022beryllium}
    \item p \& He: AMS-02~\cite{AMS:2021nhj}, DAMPE~\cite{DAMPE:2019gys, Alemanno:2021gpb}
\end{itemize}
We directly use the measurements from AMS-02 and DAMPE without introducing nuisance parameters, such as energy scale calibration, because their data is compatible within their respective error bars. To determine the total errors, we combine the systematic and statistical errors in quadrature.
The best-fit propagation parameters are listed in Table~\ref{table: propagation}, while the best-fit injection parameters are listed in Table~\ref{table: inj}. The comparison between the best-fit model results and the data of the proton and helium spectra and $\ce{^{10}Be}$/$\ce{^{9}Be}$ ratio are shown in Figs.~\ref{fig:proton} and~\ref{fig:10be9}.

\begin{table}[h]
\centering
\caption{Propagation parameters.}
\begin{tabular}{ccccc}
\hline
\hline
$D_0$ & $\delta$ & $z_h$ & $v_A$ & $\eta$\\
($10^{28} \text{cm}^2 \text{s}^{-1}$) & & (kpc) & (km s$^{-1}$) & \\
\hline
5.6 & 0.50 & 5.674 & 19.0 & -0.75  \\
\hline
\label{table: propagation}
\end{tabular}
\end{table}

\begin{table}[h]
\centering
\caption{Injection parameters.}
\begin{tabular}{cccc}
\hline
\hline
 & p & He & C \& O \\
\hline
$\nu_0$ & 2.10 & 2.03 & 1.18  \\
$\nu_1$ & 2.34 & 2.30 & 2.32 \\
$\nu_2$ & 2.20 & 2.15 & 2.13 \\
$R_{\text{br}0}$(GV) & 9.0 & 1.0 & 2.1 \\
$R_{\text{br}1}$(GV) & 200 & 150 & 200/300 \\
\hline
\label{table: inj}
\end{tabular}
\end{table}

\section{Cross sections for $^4$He target channels\label{sec: heXS}}
\begin{table*}[t]
\centering
\captionsetup{justification=raggedright}
\caption{Production cross sections of deuterons for a 2.1 GeV/n $\ce{^{16}O}$ beam and a $\ce{^{12}C}$ beam upon different targets~\cite{osti_971496}.
Additionally, the nuclear charge radii of these targets given by Ref.~\cite{Angeli:2013epw} are listed.}
\begin{tabular}{cccccccc}
\hline
\hline
 & H & Be & C & Al & Cu & Ag & Pb \\
\hline
$\ce{^{16}O}$ & 152$\pm$23 & 417$\pm$37 & 406$\pm$36 & - & 682$\pm$73 & 752$\pm$97 & 945$\pm$154  \\
$\ce{^{12}C}$ & 105$\pm$15 & 329$\pm$30 & 314$\pm$28 & 411$\pm$37 & 543$\pm$58 & 690$\pm$78 & 715$\pm$124  \\
$R$ (fm) & 0.878 & 2.646 & 2.470 & 3.061 & 3.9 & 4.564 & 5.51  \\
\hline
\label{table: radii}
\end{tabular}
\end{table*}

The deuteron fragmentation cross sections of nuclei with $A > 4$ colliding with a target proton can be calculated through the factorization scheme provided by CDMP12~\cite{Coste:2011jc}, utilizing a compilation of available cross-section data. The ISM is predominantly composed of $\ce{^1H}$ and $\ce{^4He}$ in a ratio of 9:1. The contributions of nuclei with $A > 2$ incident upon helium targets shall be considered, as they account for more than 20\% of the total deuteron fluxes. 

Currently, there has been no direct measurement of the deuteron cross section for $\ce{^4He}$-target channels.  Our approach involves a two-step analysis to determine these cross sections. Firstly, we calculate some deuteron cross sections for $\ce{^4He}$-target channels with specific projectiles, based on data from other targets. Subsequently, we extrapolate deuteron cross-sections for other $\ce{^4He}$-target channels through interpolation.  
The available data for the first step are as follows.

\begin{enumerate}
 \item ~[Li75]: Lindstrom's group measured the deuteron cross section  using a 2.1 GeV/n $\ce{^{16}O}$ beam, along with 1.05 GeV/n and 2.1 GeV/n $\ce{^{12}C}$ beams upon Be, $\ce{CH_2}$, C, Al, Cu, Ag and Pb targets ~\cite{osti_971496}. The H-target data were obtained by C-$\ce{CH_2}$ subtraction.
Their measurements focused on deuterons associated with projectile fragments rather than target fragments. By extrapolating these measurements to the $\ce{^4He}$-target case, we can infer the cross sections for $\ce{^{16}O}+\ce{^4He}\to\ce{D}$ and $\ce{^{12}C}+\ce{^4He}\to\ce{D}$.

\item ~[Ab81]: The SKM-200 Collaboration measured the deuteron cross section using a 4.5 (GeV/c)/n $\ce{^4He}$ beam upon Li, C, Al, Cu, and Pb targets~\cite{SKM-200:1980bbu}. The deuterons were observed within the momentum range of 6.5 GeV/c and 10.8 GeV/c, indicating that they exclusively originated from the projectile $\ce{^4He}$, rather than from the heavier target nuclei. By extrapolating these findings to the \ce4He $\ce{^4He}$-target case, we can infer the cross section for $\ce{^4He}+\ce{^4He}\to\ce{D}$.
\end{enumerate}

According to the analysis of [Li75], the production cross sections can be factored as $\sigma=\gamma^F_P\gamma_T$, where $\gamma^F_P$ depends on the projectile P and fragment F, while $\gamma_T$ mainly depends on the target T.
The study revealed that $\gamma_T$ could be effectively modeled by $\gamma_T=A_T^{1/4}$ or $\gamma_T\propto(A_B^{1/3}+A_T^{1/3}-0.6)$.  However, the inability to account for
$\gamma_H$ and $\gamma_{Be}$ based on the available data suggests that a more precise fit can be achieved by utilizing the nuclear radius.

By analyzing the root-mean-square nuclear charge radii from Ref.\cite{Angeli:2013epw}, we find that the factor $\gamma_T$ for the deuteron cross section is proportional to the radius of the target nucleon, as listed in Table~\ref{table: radii}. 
The relation $\gamma_T\propto R_T$ enables us to interpolate the unmeasured $\ce{^4He}$-target cross sections.
Using the radius of $\ce{^4He}$ as 1.6755 fm, we derive the cross sections of $\ce{^{16}O}+\ce{^4He}\to\ce{D}$ and $\ce{^{12}C}+\ce{^4He}\to\ce{D}$ as 277.0~mb and 219.7~mb, respectively, with 10\% uncertainties.

\begin{table}[h]
\centering
\captionsetup{justification=raggedright}
\caption{Production cross sections of deuterons for a 4.5 (GeV/c)/n $\ce{^{4}He}$ upon different targets~\cite{SKM-200:1980bbu}.
Additionally, the nuclear charge radii of these targets given by Ref.~\cite{Angeli:2013epw} are listed.}
\begin{tabular}{ccccc}
\hline
\hline
 & Li & C & Al & Cu \\
\hline
$\ce{^{4}He}$ & 84$\pm$15 & 91$\pm$27 & 113$\pm$38 & 159$\pm$45  \\
$R$~(fm) & 2.444 & 2.470 & 3.061 & 3.9 \\
\hline
\label{table: radii2}
\end{tabular}
\end{table}

In Table~\ref{table: radii2} we list the measurements from [Ab81] with charge radii of the targets.
We assume that the relation $\gamma_T\propto R_T$ is valid for the helium projectile, and derive the cross section of $\ce{^4He}+\ce{^4He}\to\ce{D}$ as 60.5 mb with a 30\% uncertainty.
Additionally, we calculate the cross section of $\ce{^4He}+p\to\ce{D}$ as 31.7 mb, which aligns closely with the value of $\sim 29$ mb reported in GALPROP v57 \cite{Porter:2021tlr} based on available measurements.

Based on the derived cross sections for the three $\ce{^4He}$-target channels, we can calculate the cross section for other $\ce{^4He}$-target channels through interpolation.
We assume that the deuteron cross sections of nuclei incident upon the $\ce{^4He}$-target as
\begin{equation}
\begin{aligned}\label{hexs factorization}
\sigma^{\mathrm{P}+\ce{^{4}He} \rightarrow \ce{D}}\left(E_{k / n}, A_{\mathrm{P}}\right)=&\gamma_{\mathrm{P}\ce{^{4}He}}^{\ce{D}} \times \sigma_{\text {breakup }}^{\ce{^{4}He}+p\to \ce{^{3}He}}\left(E_{k / n}\right)\;.
\end{aligned}
\end{equation}
By fitting the derived cross section values, we obtain $\gamma_{\mathrm{P}\ce{^{4}He}}^{\ce{D}}=\rm 0.6A_{\mathrm{P}}^{1.05}$, which is two times of the value in Eq.~(\ref{gamma D}). This result implies that the reaction of $\ce{P}+\ce{^4He}\to\ce{D}$ can be equivalent to the reaction of the projectile upon two protons to produce deuterons.

\end{document}